\documentclass[aps,twocolumn,showpacs,floatfix]{revtex4}

\usepackage{dcolumn,multirow}
\usepackage{graphicx}
\usepackage{amsfonts,mathtools,amsmath,bm,amssymb}
\usepackage{blkarray}
\usepackage{comment}
\usepackage[usenames]{color}
\usepackage[flushleft]{threeparttable}
\usepackage{titlesec}
\usepackage{etoolbox} 
\usepackage{lipsum} 
\usepackage[capitalize]{cleveref}
\usepackage{makecell}
\usepackage{ulem}
\usepackage{float}

\usepackage{color}

\begin{document}


\title {Effect of Off-Diagonal Elements in Wannier Hamiltonian on DFT+DMFT for low-symmetry material:
Study of Li$_2$MnO$_3$}


\author{Alex Taekyung Lee}
\affiliation{Department of Chemical engineering, University of Illinois at Chicago, Chicago, IL 60608, USA}
\affiliation{Materials Science Division, Argonne National laboratory, Lemont, IL 60439, USA}

\author{Hyowon Park}
\affiliation{Department of Physics, University of Illinois at Chicago, Chicago, IL 60608, USA}
\affiliation{Materials Science Division, Argonne National laboratory, Lemont, IL 60439, USA}

\author{Anh T. Ngo} \thanks{\href{mailto:anhngo@uic.edu}{anhngo@uic.edu}}
\affiliation{Department of Chemical engineering, University of Illinois at Chicago, Chicago, IL 60608, USA }
\affiliation{Materials Science Division, Argonne National laboratory, Lemont, IL 60439, USA }

\date{\today }

\begin{abstract}

We study the effect of the off-diagonal elements of the Wannier Hamiltonian  
on the electronic structure of low-symmetry material Li$_2$MnO$_3$ ($C2/m$), 
using dynamical mean field theory calculations with continuous-time 
Quantum Monte Carlo impurity solver. 
Presence of significant off-diagonal elements leads to a pronounced suppression of the energy gap.
The off-diagonal elements are largest when the Wannier projection is used 
based on the global coordinate, and they remain substantial even with the projection 
using the local coordinate close to the direction of Mn-O bonds.
We show that the energy gap is enhanced by the diagonalization of the Mn $d$ block 
in the full $p$-$d$ Hamiltonian, with  applying unitary rotation matrix.
Additionally, the inclusion of a small double counting energy is crucial for 
achieving the experimental gap by reducing $p$-$d$ hybridization. 
Furthermore, we establish the efficiency of a low-energy ($d$-only basis) model for 
studying the electronic structure of Li$_2$MnO$_3$, as the Wannier basis represents 
a hybridized state of Mn $d$ and O $p$ orbitals. 
These findings suggest an appropriate approach for investigating low-symmetry materials 
using the DFT+DMFT method. 
We also find that the antiferromagnetic ground state $\Gamma_{2u}$ is stable with $U \leq 2$  eV
within density functional theory+$U$ calculations, which is much smaller than widely used $U$=5 eV.

\end{abstract}

\pacs{ }

\maketitle

\section{Introduction}
Beyond density functional theory (DFT), dynamical mean field theory (DMFT) is 
one of the most successful method 
which account for the many-body correlation \cite{KotliarRMP2006,PRBHaule2007}.
In DMFT, the lattice problem is mapped onto an effective impurity problem, and solving this impurity model 
accurately is crucial for obtaining reliable results. 
Various impurity solvers have been developed \cite{GeorgesRMP1996}, including
Hirsh-Fye Quanum Monte Carlo (QMC) Method \cite{HirschPRL1986}, 
continuous time quantum Monte Carlo (CTQMC) \cite{PRBHaule2007,Gull_2008,GullRMP2011},
exact diagonalization (ED) \cite{CaffarelPRL1994,GeorgesRMP1996},
numerical renormalization group (NRG) \cite{BullaPRL1999}, 
and density matrix renormalization group (DMRG) \cite{WhitePRL1992} methods.

Each method has advantages and disadvantages, and there is no method that 
provides the solution both accurately and efficiently, for all regimes of parameters. 
For example, ED solver does not have sign problem, but ED is computationally 
challenging if the number of bath sites needs to be increased \cite{GeorgesRMP1996}.
Recently, CTQMC method has gained popularity as a solver for numerous DMFT applications,
because it provides accurate solution over a wide range of parameter values.
In principle, CTQMC can treat the general hybridization matrix, but the large
off-diagonal terms sometimes produce a sign problem.
The off-diagonal elements of the Hamiltonian (or density matrix) are 
usually small in the high-symmetry structure, and thus it has been neglected.
While treating non-diagonal Weiss fields within CTQMC presents no inherent conceptual challenge, 
several studies have aimed to minimize the off-diagonal terms through basis transformation, 
especially in the presence of significant spin-orbit coupling \cite{SatoPRB2015, AaramPRL2017}.

In the DFT+DMFT procedure \cite{DMFTwDFT}, the Hamiltonian with a localized basis,
such as maximally localized Wannier functions \cite{MLWF}, 
is obtained from the DFT calculations.
When the crystal symmetry of the transition metal (TM) ion is low, 
substantial off-diagonal elements in the Wannier Hamiltonian also emerges
due to the significant mixing between $d$ basis.
For example, the non-negligible off-diagonal terms have been reported in 
titante \cite{KrabergerPRB2017} or vanadate \cite{BeckPRB2018}.

\begin{figure}
\begin{center}
\includegraphics[width=0.48\textwidth, angle=0]{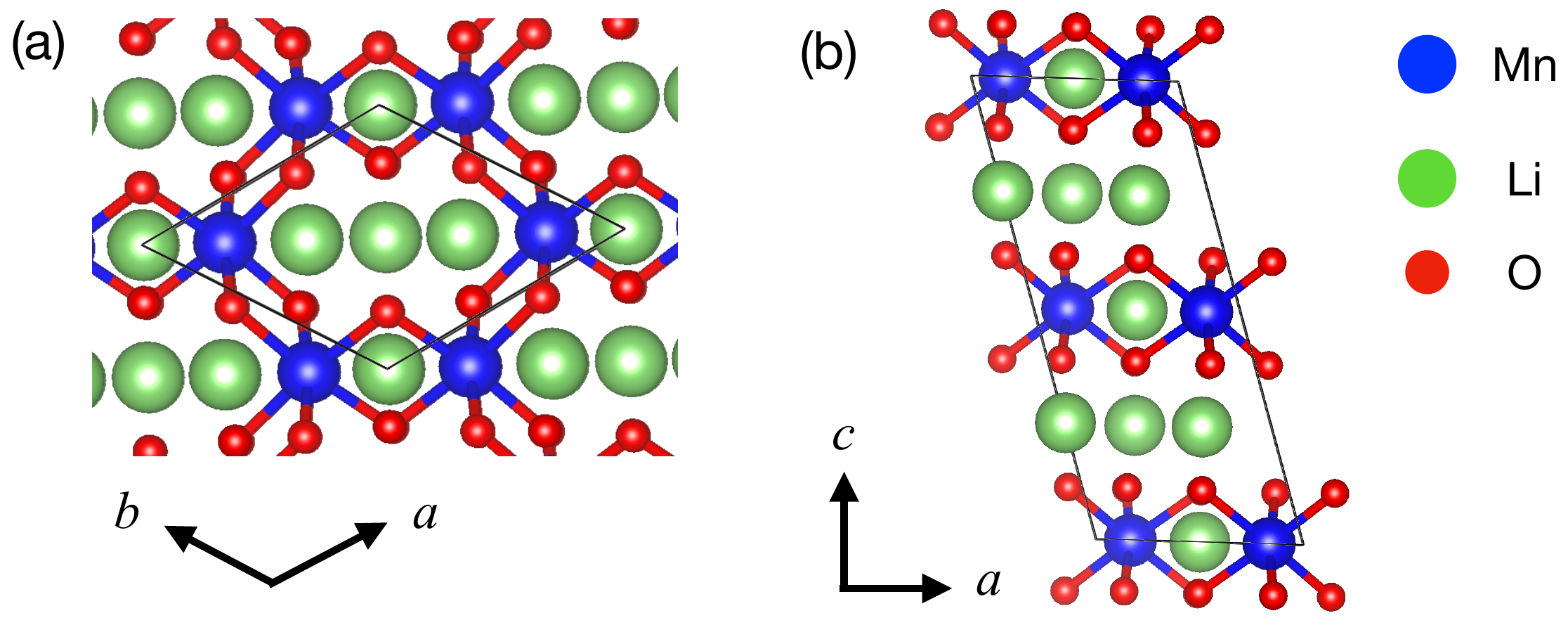}
\includegraphics[width=0.48\textwidth, angle=0]{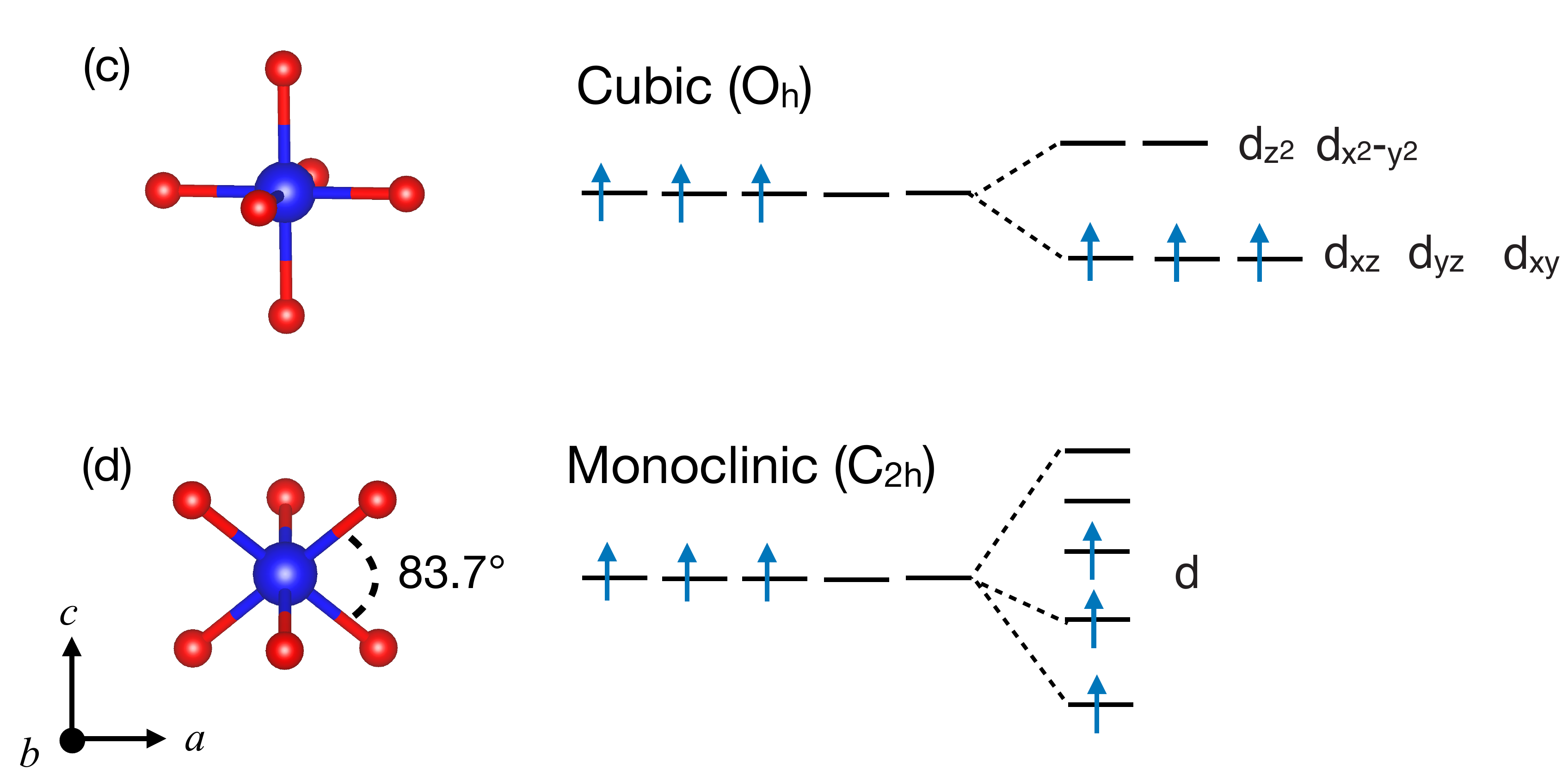}
\caption{ (a) top and (b)side view for the atomic structure of Li$_2$MnO$_3$. 
Li atoms are located between MnO$_3$ layers and the center of the hexagon.
Schematic band splitting based on the (c) cubic ($O_h$) and (d) monoclinic ($C_{2h}$) crystal fields. 
Vectors $\mathbf{a}$, $\mathbf{b}$, and $\mathbf{c}$ indicate the global coordinates.
Note that Mn-O bonds are the global axes are not parallel in the monoclinic phase.
}
\label{atm_str}
\end{center}
\end{figure}

Li$_2$MnO$_3$, one of the potential candidates for 
next generation cathode material due to the high voltage 
(4.4$-$5 V) and the low cost \cite{Yabuuchi2011,SeoNchem2016},
shows a good scenario for studying the impact of off-diagonal elements 
becasue of its low crystal symmetry.
Since Li$_2$MnO$_3$ has monoclinic structure with $C/2m$ space group (No. 12),
much lower symmetry than titanates and vanadates ($Pbnm$, No. 62), 
even larger off-diagonal terms are expected. 
Since its six Mn-O bonds are neither parallel nor perpendicular 
(see Fig. \ref{atm_str}), a substantial degree of mixing between 
the $d$ orbitals is expected.

Li$_2$MnO$_3$ is an insulator with an experimental band gap ranging from 
2.1 \cite{Tamilarasan2015} to 2.17 eV \cite{singhSNAS2020}. 
The Mn ion in Li$_2$MnO$_3$ exhibits a 4+ charge state ($d^3$) and 
a high-spin configuration ($S=3/2$), resulting in a magnetic moment of 
2.3$-$2.7 $\mu_B$ \cite{LeeJPCM2012, StrobelJSSC1988}.
In the high-spin state, the three $d$ electrons fully occupy the spin-up $t_{2g}$ band, 
leading to a non-zero gap for Mn$^{4+}$ due to crystal field splitting and Hund coupling. 
At low temperatures, Li$_2$MnO$_3$ exhibits an antiferromagnetic phase with 
a N\'eel temperature of $T_N$= 36$-$36.5 K \cite{StrobelJSSC1988, LeeJPCM2012}.
Early studies of Li$_2$MnO$_3$ reported the antiferromagnetic ground state to be $\Gamma{3g}$ 
with a magnetic propagation vector of $Q_m = (0, 0, 0.5)$ \cite{StrobelJSSC1988}. 
However, recent studies have shown that the $\Gamma_{2u}$ model with $Q_m = (0, 0, 0.5)$ 
provides the best agreement with neutron diffraction data, while the $\Gamma_{3g}$ model 
does not match the full refinement results \cite{LeeJPCM2012}.

Numerous studies based on density functional theory (DFT) have investigated 
the electronic and magnetic structures of Li$_2$MnO$_3$ 
\cite{XiaoCM2012, SeoNchem2016, WangJMCA2017, HungruCM2016, ChenJACS2019, HikimaJACS2022}. 
However, a comprehensive understanding of these structures is still lacking. 
The band gaps have been predicted through various approaches such as DFT+$U$ 
\cite{XiaoCM2012, WangJMCA2017}, GW calculations \cite{SeoNchem2016}, or 
hybrid functionals \cite{SeoNchem2016, HungruCM2016}. 
Notably, previous studies have primarily focused on nonmagnetic \cite{SeoNchem2016, HikimaJACS2022} 
or ferromagnetic \cite{XiaoCM2012, HungruCM2016, WangJMCA2017} configurations, 
neglecting the magnetic ground state.

Understanding the paramagnetic phase of Li$_2$MnO$_3$ holds great significance, 
particularly considering the operational temperature range of room temperature and 
the relatively low N\'eel temperature ($T_N$) of Li$_2$MnO$_3$ at 36 K \cite{LeeJPCM2012}. 
However, conventional nonmagnetic calculations fail to accurately describe the 
paramagnetic spin order due to the absence of a local spin in the nonmagnetic phase, 
while the paramagnetic phase exhibits an averaged spin of zero due to the fluctuations of the local spin.
To investigate the paramagnetic phase of Li$_2$MnO$_3$, 
recently developed DFT-based methods 
\cite{GambinoPRB2022, TrimarchiPRB2018, DU2022100805} 
or many-body techniques such as dynamical mean-field theory (DMFT) are required.

Other transition metal oxides for the cathode Li-ion batteries
such as LiCoO$_2$ \cite{IsaacsPRB2020} and LiNiO$_2$ \cite{KorotinPRB2019}, 
which have higher symmetry ($R\bar{3}m$, No. 166), are also studied within DFT+DMFT. 
Interestingly, it has been shown that the electronic structure and the size of the energy gap in
LiNiO$_2$ strongly rely on the choice of the different Wannier basis \cite{KorotinPRB2019}.
Since the symmetry of Li$_2$MnO$_3$ ($C2/m$) is lower than the symmetry of LiNiO$_2$ 
($R\bar{3}m$), the distortion of TM-O octahedron is larger in Li$_2$MnO$_3$.
Consequently, the off-diagonal elements in Li$_2$MnO$_3$ become more pronounced and play a crucial role. 
Therefore, a systematic study to examine the impact of these off-diagonal elements 
and the selection of various Wannier bases within the DFT+DMFT framework is needed for Li$_2$MnO$_3$.

In this study, we explore the influence of off-diagonal elements in the Wannier Hamiltonian 
on the electronic structure of Li$_2$MnO$_3$ using DFT+DMFT.  
The low symmetry of Li$_2$MnO$_3$ ($C2/m$) results in significant off-diagonal terms, 
leading to a suppressed energy gap within DMFT using $U$=5 eV. 
Although a Wannier basis aligned with the local MnO$_6$ coordinate reduces off-diagonal elements, 
they remain large, resulting in a suppressed gap due to the low point group symmetry. 
We apply a unitary rotation matrix to diagonalize the block diagonal part of the Mn $d$ Hamiltonian 
to mitigate the effect. 
However, the resulting gap remains smaller than the experimental value, 
and increasing $U$ does not resolve the discrepancy. 
We use a small double counting energy to obtain the experimental band gap, 
considering the influence of $p$-$d$ covalency. 
Additionally, we find that a minimal $d$-only Wannier basis efficiently captures the 
electronic structure by encompassing the hybridized states of Mn $d$ and O $p$ orbitals.

\section{Methods}

\subsection{DFT+DMFT}
\label{sec:method_dmft}

\begin{figure}
\begin{center}
\includegraphics[width=0.45\textwidth, angle=0]{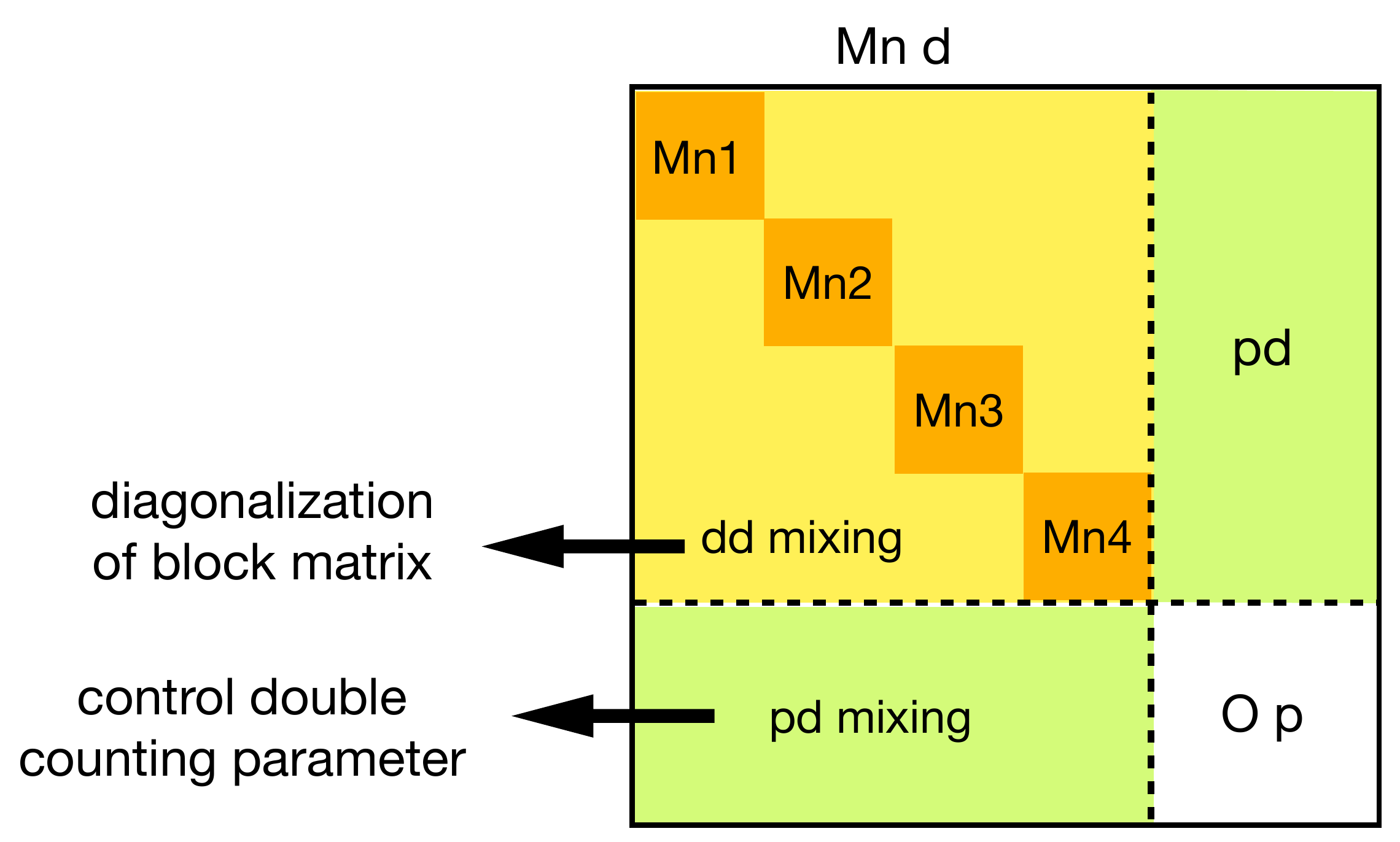}
\caption{  Schematic diagram of the Wannier Hamiltonian matrix for Mn $d$ and O $p$ orbitals. 
We diagonalize the diagonal blocks of each Mn $d$ states (orange boxes), 
and control $p-d$ hybridization by changing double counting parameter.
}
\label{matrix_diag}
\end{center}
\end{figure}

We employ the non-charge-self-consistent DFT+DMFT method \cite{PRB2020Hyowon,DMFTwDFT} 
for relaxed structures obtained from DFT calculations. 
For DFT calculation, we use the projector augmented wave (PAW) method \cite{PAW} 
and the revised version of the generalized gradient
approximation (GGA) proposed by Perdew {\it et al.} (PBEsol) \cite{PBEsol} 
as implemented in the VASP software \cite{VASP}.
Spin-independent version of the exchange correlation functional are employed.
A plane wave basis with a kinetic energy cutoff of 500 eV is used. 
We used 24 atom unit cells (i.e., $1\times 1\times 2$ unit cells), which contains 4 Mn atoms, 
and $\Gamma$-centered \textbf{k}-point meshes of size  10$\times$10$\times$5.
Atomic positions within the unit cells were relaxed until the residual forces were less than 
0.01 eV/\AA, and the stress was relaxed  below 0.02 kB.

We solve the many-body problem on the manifold of both Mn 3$d$+O $p$ Wannier orbitals 
and Mn 3$d$-only orbitals.
The DFT+DMFT calculation has the following steps. 
First, we solve the non-spin-polarized Kohn-Sham (KS) equation within DFT using VASP.  
Second, we construct a localized-basis Hamiltonian for the Mn 3$d$ bands by generating maximally 
localized Wannier functions (MLWFs) \cite{MLWF} for the nonmagnetic DFT band structure.  
The energy window employed ranges from $E_F?9.0$ eV to $E_F+5.5$ eV for the $pd$ basis,
and $E_F?2.0$ eV to $E_F+4.0$ eV for the $d$-only basis.
Finally, we solve the DMFT self-consistent equations for the correlated subspace of 
Mn 3$d$ and O $p$ Wannier orbitals (or only Mn 3$d$ orbitals) using the continuous time 
quantum Monte Carlo (CTQMC) \cite{PRBHaule2007,ctqmcRMP2011} impurity solver.

Coulomb interaction element $U_{m_4 m_3 m_2 m_1}$ in the Slater Hamiltonian 
using spherical harmonics function $Y_{lm}$ is given by
\begin{equation}
\begin{split}
U_{m_4 m_3 m_2 m_1} = &\sum_{k} \frac{4\pi}{2k+1} F^{k}_l 
\left< Y_{lm_4} \left| Y_{k,m_4 - m_1} \right| Y_{lm_1}\right> \\
& \times \left< Y_{lm_3} \left| Y^*_{k,m_2 - m_3} \right| Y_{lm_2}\right>
\end{split}
\end{equation}
where $F^k$ are Slater integrals. 
Both Hubbard $U$ and Hund's couplings $J$ are parameterized by the  
Slater integrals, using $U = F^0$ and $J = (F^2 + F^4)/14$. 
We consider full Coulomb interaction including density-density 
(where $m_1 = m_4$ and $m_2 = m_3$),
spin-flip, and pair-hopping terms. 
To study the effect of the off-diagonal elements of the Coulomb interaction matrix
on the electronic structure, we also use only density-density interaction term and compare 
with the full Coulomb interaction calculations.
We note that results using density-density approximation and full Coulomb interaction are 
qualitatively same (see Appendix \ref{appendix:coulomb}).
For the $pd$ basis Hamiltonian, we used $U$ values of 5 and 7 eV and $J$ of 0.9 eV.
We used electronic temperatures of 300 K to study the temperature effect on the spectral function.

Within DFT+DMFT framework \cite{DMFTwDFT}, the self-energy convergence is achieved when 
$\Sigma^{\textrm{loc}} ( i \omega) = \Sigma^{\textrm{imp}} (i \omega)$, where 
$\Sigma^{\textrm{loc}} ( i \omega)$ and $\Sigma^{\textrm{imp}} (i \omega)$
are local and lattice self-energies, respectively, and $i\omega$ is imaginary frequency.
$\Sigma$ is approximated as a local quantity in the correlated subspace.
DFT+DMFT total energy is given by 
\begin{equation}
 \label{E_total1}
\begin{split}
    E^{\textrm{TOT}} =& E^{\textrm{DFT}} (\rho) 
    + \sum_{m, \mathbf{k}} \epsilon_{m}(\mathbf{k}) \cdot \big [ n_{mm}(\mathbf{k}) - f_{m}(\mathbf{k}) \big ] \\
    &+ E^{\textrm{POT}} - E^{\textrm{DC}},
\end{split}
\end{equation}
where $E^{\textrm{DFT}}$ is the total energy from non spin-polarized DFT, and 
$\epsilon_{m}(\mathbf{k})$ are the DFT eigenvalues.
$n_{mm}(\mathbf{k})$ and $f_{m}(\mathbf{k})$ are the diagonal DMFT occupancy matrix element 
and Fermi function, respectively, for $m$th KS band and momentum $\mathbf{k}$.
The potential energy $E^{\textrm{POT}}$ is calculated by using Galitskii-Migdal formula \cite{Galitskii}: 
\begin{equation}
    E^{\textrm{POT}} = \frac{1}{2}  \sum_{\omega} \big [ \Sigma^{\textrm{loc}}(i\omega) \cdot G^{\textrm{loc}} (i\omega)].
\end{equation}
Here, the local Green's function is simplified by 
$G^{\textrm{loc}} (i\omega) = \sum_{\mathbf{k}} G^{\textrm{loc}} (\mathbf{k},i\omega)$.

To obtain the spectral function, the maximum entropy method is used for the analytic continuation.
Spectral function $A(\omega)$ is given by
\begin{equation}
    A(\omega) = -\frac{1}{\pi} \textrm{Im} \Big [ \sum_{\mathbf{k}} G^{\textrm{loc}}(\mathbf{k}, \omega)\Big].
\end{equation}
%

\subsubsection{Different Wannier basis and Diagonalization of the Wannier Hamiltonian}

When investigating the electronic structure of Li$_2$MnO$_3$ using the DMFT method, 
it is crucial to address two key issues:
(i) the presence of significant off-diagonal terms in the Wannier Hamiltonian 
(yellow region in Fig. \ref{matrix_diag}), and
(ii) the significance of $p-d$ hybridization  
(green region in Fig. \ref{matrix_diag}).
To overcome these challenges, we employed two strategies:
(i) diagonalizing the block $d$ Hamiltonian, and
(ii) utilizing different values of double counting parameters, 
as summarized in Figure \ref{matrix_diag}.

Substantial off-diagonal elements in the Hamiltonian may induce notable 
inaccuracies in DMFT calculations. 
The non-parallel alignment of the cartesian axes of the Wannier orbitals 
and the directions of the Mn-O bonds arises from the $C_{2h}$ point group symmetry 
of the MnO$_6$ octahedron. 
In cases where the point group symmetry of the transition metal (TM) octahedron 
is non-cubic, such as trigonal or monoclinic, there is a substantial mixing of the 
$d$ basis ($d{xy}$, $d_{xz}$, $d_{yz}$, $d_{z^2}$, $d_{x^2-y^2}$), 
as these bases are defined within the cubic crystal field framework. 
Consequently, the off-diagonal terms of the Wannier Hamiltonian with the cubic 
$d$ orbital basis become significant, leading to errors within the DMFT calculations. 
It is noteworthy that DFT+$U$ does not encounter such issues since DFT+$U$ exhibits 
rotational invariance \cite{LDA+U1}, allowing for the inclusion of off-diagonal terms in the density matrix.

We employed three distinct approaches to investigate the impact of off-diagonal elements 
in Li$_2$MnO$_3$:
(i) utilizing the global coordinate system,
(ii) adopting the local Mn--O bond coordinate system, and
(iii) diagonalizing the Mn $d$ blocks (orange region in Fig. \ref{matrix_diag}) 
of the Hamiltonian by applying a unitary rotation matrix.
The selection of a suitable local coordinate system is not straightforward 
due to the non-perpendicular arrangement of Mn--O bonds, 
as depicted in Fig. \ref{atm_str}(d). 
To address this, we designated the longest Mn-O bond as the local $z$-axis 
and established local $x$- and $y$-axes that were perpendicular to the c-axis. 
We then minimized the displacement between the real Mn-O bond and the local $x$- and $y$-axes.

\subsubsection{Controlling $p$-$d$ Covalency through Double Counting Energy}
\label{sec:Wann_basis}

On the other hand, in cases where $p-d$ hybridization is as pronounced as in nickelates \cite{HyowonPRB2015,HyowonPRL2012}, applying nonzero $U$ to the $p-d$ Wannier Hamiltonian 
might not sufficiently capture the intrinsic physics of the $p-d$ hybridization.
To resolve this issue, we use the double counting parameter, which control the degree of $p-d$ covalency.
In term of double counting corrections for DFT+DMFT, we use a double counting energy 
($E^{\rm{DC}}$) and potential ($V^{\rm{DC}} = \partial E^{\rm{DC}} / \partial N_d$)
similar to the conventional fully localized limit \cite{AnisimovPRB1991,PRB2020Hyowon}:

\begin{equation}
E^{\rm{DC}} = \frac{U}{2} N_d  \cdot \left( N_d - 1 \right) - \frac{J}{4} N_d \cdot \left( N_d - 2 \right)\,,
\label{eq:Edc}
\end{equation}
\begin{equation}
V^{\rm{DC}}  = \frac{U}{2} \left( \overline{N}_d  - \frac{1}{2} \right)  - \frac{J}{2} \left( \overline{N}_d - 1 \right) 
\label{eq:Vdc}
\end{equation}
Here $\overline{N}_d = N_d - \alpha$, where $N_d$ is the $d$ occupancy obtained self-consistently 
at each Mn site, and $\alpha$ is double counting parameter. 
$N_d$ is computed from the local Green function $G^{\textrm{loc}}(\mathbf{k}, \mathbf{k}^{\prime},i \omega)$:
\begin{equation}
\label{Nd1}
    N_d = \sum_{a, \omega} \sum_{ \mathbf{k}, \mathbf{k}^{\prime}}  \textrm{Im} \big \{  [ \phi^a_d(\mathbf{k})]^{\ast} G^{\textrm{loc}}(\mathbf{k}, \mathbf{k}^{\prime}, i \omega) \phi^{a}_d (\mathbf{k}^{\prime}) \big \},
\end{equation}
where $\phi^a_d(\mathbf{k})$ is the normalized $d$-orbital wavefunction, 
which is transformed from the wavefunction in the real space
$\phi^a_d(\mathbf{r})$ with the center of coordinates 
$\mathbf{r}$ on a transition metal ion.
Note that $\alpha = 0$ gives the conventional fully localized limit. 
From the Eqs. \ref{eq:Edc} and \ref{eq:Vdc}, changing $\alpha$ (or $V^{\rm{DC}}$) can tune the 
$p-d$ covalency by effectively shifting the $d$ orbital level.
If $V^{\rm{DC}}$ potential is smaller than the DC potential of fully localized limit, 
it will make $d$ orbital level higher and the covalency effect weaker, with reduced $N_d$.

\subsection{DFT+$U$ and atomic structures}
\label{sec:method_dft+u}

The GGA+$U$ scheme within the rotationally invariant formalism together with the fully localized
limit double-counting formula \cite{LDA+U1} is used to study the effect of electron interactions.  
We considered three different Hund's parameter ($J$) values, $J$= 0, 0.5, and 0.9.
It is worth mentioning that when $J=0$, the exchange interaction is already included 
in the spin-dependent DFT exchange-correlation potential, 
as demonstrated in a previous study \cite{HyowonPRB2015}. 
While the consideration of the full Coulomb vertex is crucial in certain systems 
\cite{BultmarkPRB2009}, the electronic structures of Li$_2$MnO$_3$ obtained 
with non-zero $J$ values are qualitatively similar to those obtained with $J=0$, 
as detailed in Appendix \ref{appendix:dftu+J}. 
The band gap is robust on $J$, while the larger values of $J$ lead to an 
increased critical $U$ value for the magnetic transition.
Therefore, unless otherwise specified, we adopt $J=0$ throughout the manuscript.

Projected density of states (PDOS) are obtained by the spherical harmonic projections 
inside spheres around each atom.
Wigner-Seitz radii of 1.323 \AA~ were used for the projection of Mn atoms, 
respectively, as implemented in the VASP-PAW pseudopotential.

Both spin-independent and spin-dependent versions of the exchange correlation 
functional are employed in the DFT+$U$ calculations. 
The structural relaxations, including the relaxation of internal forces and stress, 
were performed independently for each magnetic configuration, 
as well as for each value of $U$ and $J$.

\section{ Results and discussion}

At low temperature, Li$_2$MnO$_3$ is antiferromagnetic with $T_N$= 36K \cite{LeeJPCM2012}.
There are many DFT+$U$ studies of Li$_2$MnO$_3$, using $U$(Mn) = 5eV
\cite{XiaoCM2012,WangJMCA2017,HungruCM2016,ChenJACS2019,HikimaJACS2022},
but the previous studies only considered nonmagnetic or ferromagnetic configuration.
Therefore, in Secion \ref{sec:dft+u}, we systematically study the electronic structures 
and magnetic stabilities of Li$_2$MnO$_3$ using DFT and DFT+$U$.

In addition, Li$_2$MnO$_3$ has paramagnetic spin configuration, which is not studied yet.
 In Section \ref{sec:dft+dmft}, we delve into the electronic structure of Li$_2$MnO$_3$ 
 within the DFT+DMFT method, enabling a systematic examination of the effects of 
 off-diagonal terms in the Wannier Hamiltonian.

\begin{figure}
\begin{center}
\includegraphics[width=0.45\textwidth, angle=0]{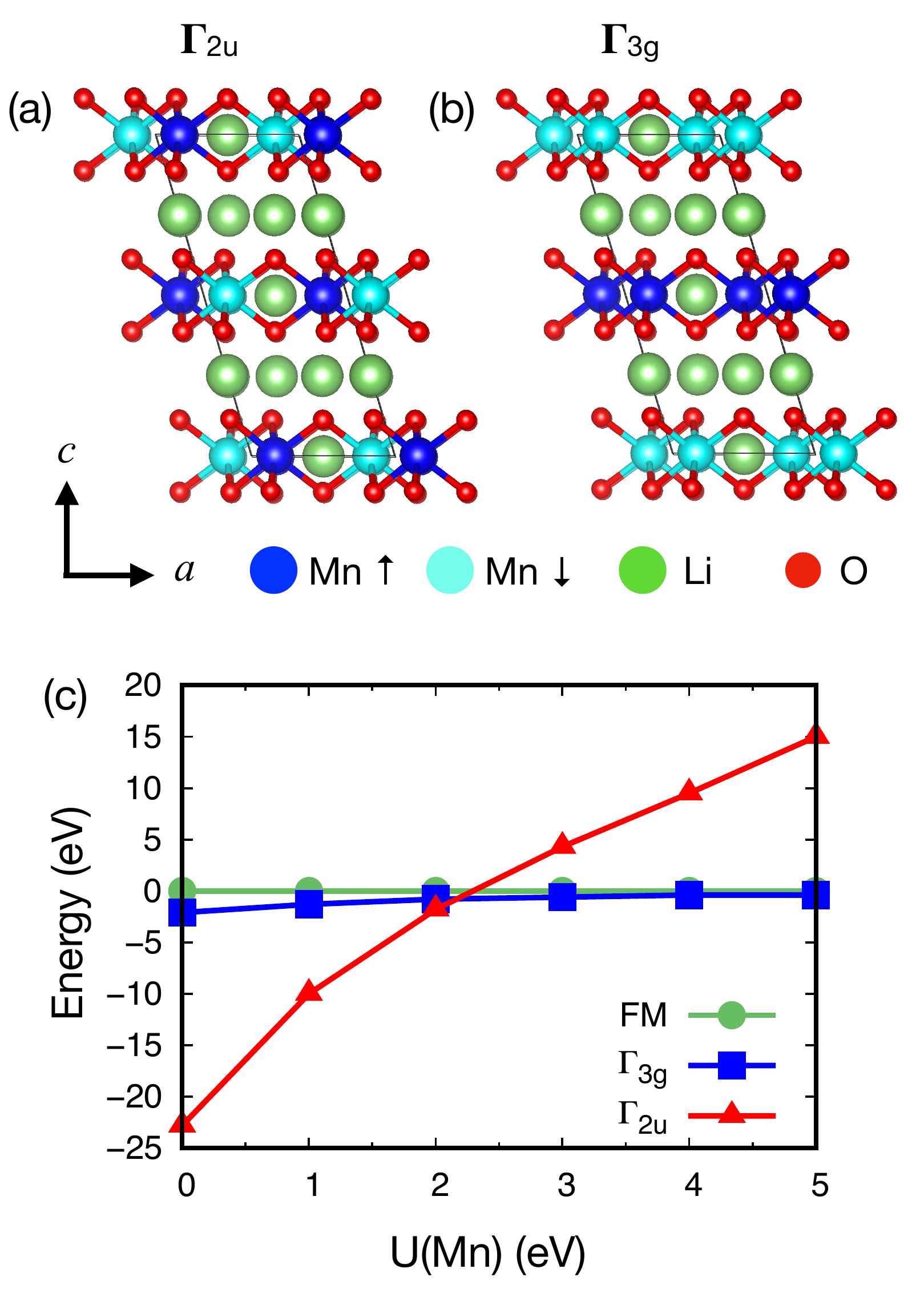}
\caption{  Possible magnetic phases of Li$_2$MnO$_3$ at low temperature, (a) $\Gamma_{2u}$ and (b) $\Gamma_{3g}$.
Mn ions with spin-up and spin-down are represented by different colors (blue and skyblue).
(c) the relative energies of the ferromagnetic, $\Gamma_{2u}$, and $\Gamma_{3g}$ phases, as a function of $U$(Mn).
Energy of the ferromagnetic phase is set to be zero. }
\label{mag_str}
\end{center}
\end{figure}

\subsection{DFT+U}
\label{sec:dft+u}

We begin the discussion by studying the magnetic phase of Li$_2$MnO$_3$.
Li$_2$MnO$_3$ is antiferromagnetic with Neel temperature of $T_N = 36$ K, 
and the magnetic propagation vector is $Q_m = (0, 0, 0.5)$ \cite{StrobelJSSC1988}.
Earlier study suggested that $\Gamma_{3g}$ is the ground state \cite{StrobelJSSC1988}.
However, recent neutron diffraction study for both powder and single crystal 
showed that the ground state is $\Gamma_{2u}$,
and $\Gamma_{3g}$ model does not match with the experiment from the full refinement
 \cite{LeeJPCM2012}.
Magnetic moment of Mn is 2.3$-$2.7 $\mu_B$ \cite{LeeJPCM2012,StrobelJSSC1988}, 
indicating that Mn has high-spin state.

In Figure \ref{mag_str}, we compare the energies of the ferromagnetic (FM), 
antiferromagnetic $\Gamma_{3g}$ and $\Gamma_{2u}$ phases, as a function of $U$(Mn).
$\Gamma_{2u}$ phase is most stable if $U$(Mn) $\leq 2$, while $\Gamma_{3g}$ becomes 
more stable when $U$(Mn) $> 2$ eV, while the Mn ion always has the high-spin state.
This result shows that $U$(Mn) $\leq 2$ eV is needed to obtain the experimental ground state.
Previous linear response calculation for spinel Mn$^{4+}$ suggested that 
$U$(Mn)= 5.04 eV \cite{ZhouPRB2004}, and many DFT+$U$ studies 
\cite{XiaoCM2012,WangJMCA2017,HungruCM2016,ChenJACS2019,HikimaJACS2022}
used $U$(Mn)= 5 eV based on this study.

However, it is important to note that using $U$ values obtained from linear 
response theory within GGA+$U$ can lead to an overestimation of correlation effects
in transition metal oxides.
This is typical in DFT+$U$ since the Coulomb interaction is treated in a 
Hartree-Fock-like fashion, neglecting local correlation effects. 
For example, quantities such as bond length disproportionation can be 
overestimated using DFT+$U$, and this overestimation can be corrected 
by considering the local correlation 
effects through the use of DMFT \cite{ParkPRB2014}. 
In light of this, we suggest that for the study of electronic and magnetic 
properties of Li$_2$MnO$_3$ within DFT+$U$, a reasonable value for $U$(Mn) is 2 eV.

Interestingly, we find that the in-plane magnetic interaction in Li$_2$MnO$_3$ is 
strong and sensitive to $U$, while the out-of-plane magnetic interaction is 
weak and less affected by $U$. 
Specifically, the energy difference between the ferromagnetic (FM) and $\Gamma_{3g}$ 
antiferromagnetic phases, $E[\text{FM}] - E[\Gamma_{3g}]$, ranges from 2.1 to 0.4 eV 
as $U$ varies from 0 to 5 eV. 
This indicates a preference for antiferromagnetic ordering in the out-of-plane direction, 
but with weak magnetic stability. 
On the other hand, the energy difference between the $\Gamma_{3g}$ and $\Gamma_{2u}$ 
antiferromagnetic phases, $E[\Gamma_{3g}] - E[\Gamma_{2u}]$, ranges from 
20.7 to 15.4 meV for $U$ = 0 to 5 eV, indicating a stronger in-plane antiferromagnetic interaction. 
Li$_2$MnO$_3$ is a layered material, with Li atoms located between the MnO$_3$ layers, 
resulting in weak interlayer interactions between the MnO$_3$ layers. 
It is worth noting that the shortest Mn-Mn distances in the in-plane and out-of-plane directions 
are 2.84 and 5.01 \AA, respectively, which explains the strong in-plane and 
weak out-of-plane Mn-Mn interactions.

When $U$(Mn) $\leq 2$ eV, the energy difference between $\Gamma_{2u}$ and 
other magnetic configuration decreases as a function of $U$.
With $U$(Mn) $= 2$ eV, $\Gamma_{2u}$ is more stable than FM and $\Gamma_{3g}$ by 1.8 and 0.8 meV, 
respectively, consistent with the low N\'eel temperature $T_N = 36$ K.
Magnetic moment of Mn in $\Gamma_{2u}$ phase is 2.8 $\mu_B$, similar to the experimental value 2.3 $\mu_B$.

\begin{figure}
\begin{center}
\includegraphics[width=0.4\textwidth, angle=0]{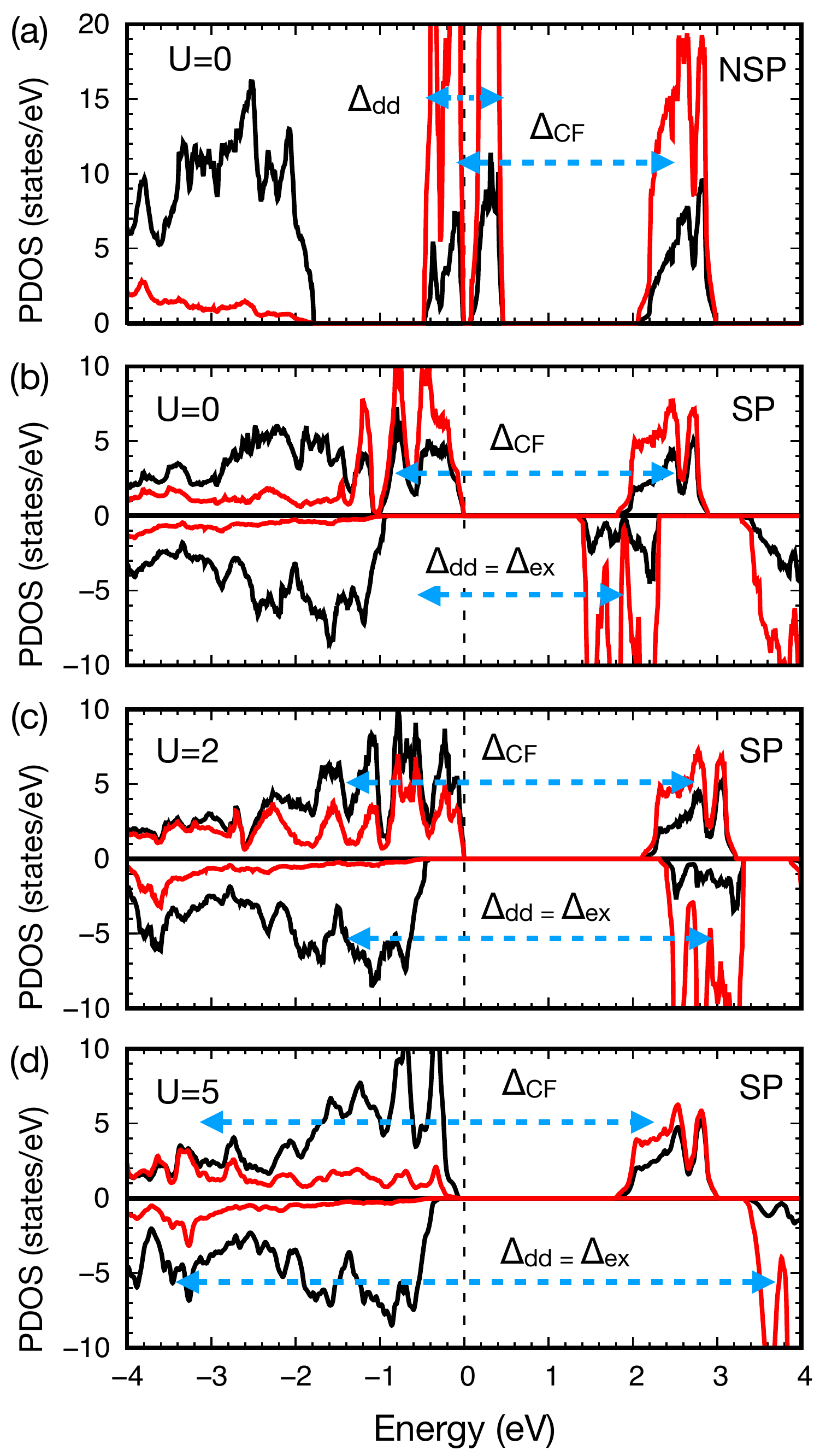}
\caption{  Projected density of states (PDOS) onto the Mn $d$ (red) and 
O $p$ (black) in Li$_2$MnO$_3$. 
(a) $U$(Mn) $= 0$ for non spin-polarized configuration,
(b) $U$(Mn) $= 0$ for spin-polarized configuration (ferromagnetic), 
(c) $U$(Mn) $= 2$ and (d) $U$(Mn) $= 5$ for spin-polarized configuration. 
The length of the arrows ($\Delta_{dd}$ and $\Delta_{CF}$) are serves as a guide for the eyes.
}
\label{dft_dos}
\end{center}
\end{figure}

Next, we study the effect of $U$ on the electronic structure and energy gap of Li$_2$MnO$_3$.
Mn ion in Li$_2$MnO$_3$ has 4+ charge state with 3 $d$ electrons, and it has high-spin state.
MnO$_6$ in $C2/m$ phase has $C_{2h}$ point group symmetry, which splits the $d$ bands into
five non-degenrate bands. 
As depicted in Fig. \ref{atm_str}, three electrons of Mn occupy the three spin-up $d$ bands
for the high-spin state.
Therefore, there are two types of splittings which determine the energy gap ($E_g$):
(i) crystal field splitting ($\Delta_{\mathrm{CF}}$) between 3 lower-energy $d$ bands 
and 2 higher-energy $d$ bands, and
(ii) $d$-$d$ splitting ($\Delta_{dd}$), which is splitting between bands with same orbital character.
Note that $\Delta_{dd}$ is equivalent to the exchange splitting ($\Delta_{\mathrm{ex}}$) 
for spin-polarized DFT. 
These splittings are also presented in the PDOS in Fig. \ref{dft_dos}.
We determine $\Delta_{\mathrm{CF}}$ and $\Delta_{dd}$ by calculating 
the energy difference between the centers of the PDOS associated with each $d$ band.

For the non-spin-polarized phase with $U$(Mn) = 0, the crystal field splitting 
($\Delta_{\mathrm{CF}}$) is around 2.5 eV, while the $d$-$d$ splitting 
($\Delta_{\mathrm{dd}}$) is only about 0.6 eV. 
As a result, the band gap ($E_g$) is almost zero, as depicted in Fig. \ref{dft_dos}(a). 
However, when spin polarization is introduced [Fig. \ref{dft_dos}(b)], 
the $d$-$d$ splitting ($\Delta_{dd}$) is significantly enhanced to approximately 2.2 eV, 
and the crystal field splitting ($\Delta_{\mathrm{CF}}$) also increases to around 3 eV. 
Since the $d$-$d$ splitting ($\Delta_{dd}$) determines the size of the energy gap ($E_g$) due to $\Delta_{dd} < \Delta_{\mathrm{CF}}$, the band gap ($E_g$) becomes 1.19 eV.

With nonzero $U$, both the $d$-$d$ splitting ($\Delta_{dd}$) 
and the crystal field splitting ($\Delta_{\mathrm{CF}}$) increase, 
as shown in Fig. \ref{dft_dos}(c). 
With $U$(Mn) = 2 eV, we obtain a band gap ($E_g$) of 2.0 eV, 
which is in good agreement with the experimental gap of 2.1$-$2.17 eV 
\cite{singhSNAS2020,Tamilarasan2015}. 
When $U$(Mn) = 5 eV, the band gap ($E_g$) is 1.9 eV, which is similar to the value 
obtained with $U$(Mn) = 2 eV. While the occupied Mn $d$ band is shifted 
to lower energies with a larger $U$ value, the energy of the valence band maximum 
is less sensitive to $U$, resulting in minimal changes to the band gap ($E_g$).

\subsection{DFT+DMFT}
\label{sec:dft+dmft}

\begin{figure}
\begin{center}
\includegraphics[width=0.4\textwidth, angle=0]{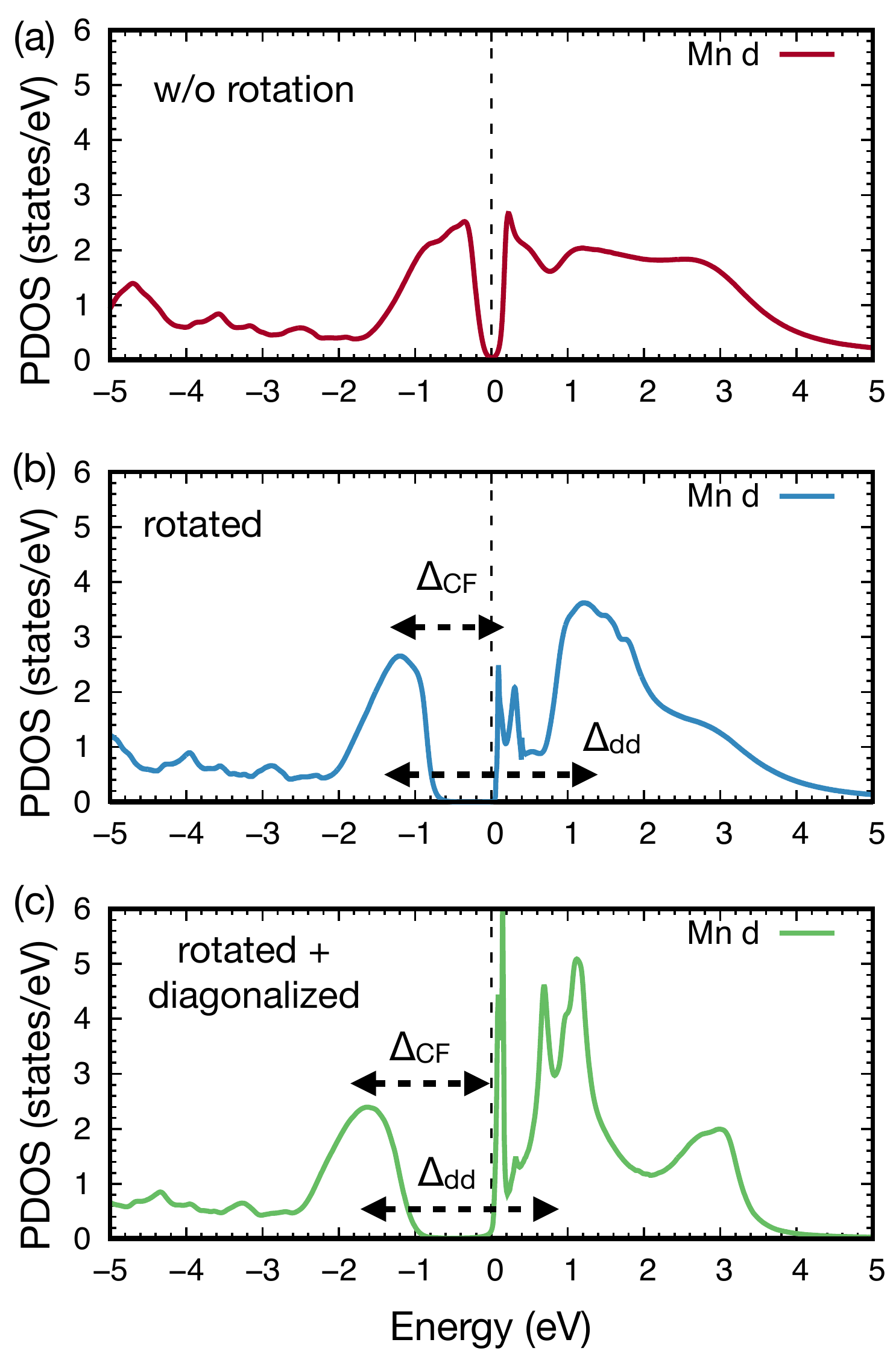}
\caption{DFT+DMFT spectral functions of Mn $d$ Wannier orbitals, with $pd$ model.
(a) global coordinate is used for Wannier projection, (b) local coordinate obtained by unitary rotation matrix
is used for Wannier projection, and (c) block diagonal part of Mn $d$ Hamiltonian is diagonalized
The calculations use $U=5$ eV,  $J=0.9$ eV, and a temperature of 300 K. 
The length of the arrows ($\Delta_{dd}$ and $\Delta_{CF}$) are serves as a guide for the eyes.s
}
\label{dmft_dos_pd}
\end{center}
\end{figure}

\begin{figure*}
\begin{center}
\includegraphics[width=0.98\textwidth, angle=0]{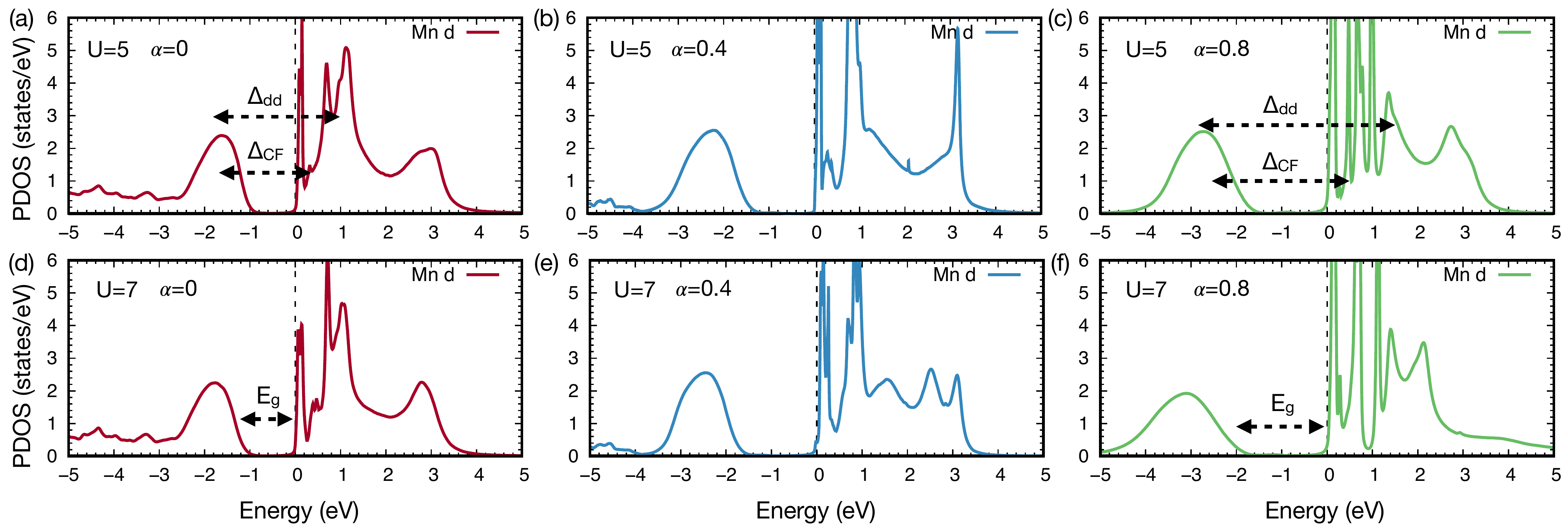}
\caption{DFT+DMFT spectral functions of Mn $d$ Wannier orbitals with different $U$ and $\alpha$.
The calculations use temperature of 300 K. 
The length of the arrows ($\Delta_{dd}$, $\Delta_{CF}$, and the energy gap $E_{g}$) 
serves as a guide for the eyes.
}
\label{dmft_dos_pd_U}
\end{center}
\end{figure*}

As discussed, the point group symmetry of MnO$_6$ in Li$_2$MnO$_3$ is $C_{2h}$, 
which is different from the cubic $O_h$ symmetry. 
This is illustrated schematically in Figs. \ref{atm_str}(c) and (d). 
In the cubic phase, the local Mn-O bonds align with the global coordinates 
based on the symmetry, resulting in a Hamiltonian based on the $d$ orbitals 
without any off-diagonal terms. 
However, in the $C2/m$ phase, the local Mn-O axes and the global axes are not parallel. 
As a result, when using global coordinates for the Wannier projection, 
significant off-diagonal terms manifest in the local on-site part of the Wannier Hamiltonian, 
as depicted in Eq. \ref{ham:noRot}.

\begin{equation}
\begin{aligned}
& \mathbf{H}_{i} = \\
& \bordermatrix{ & d_{z^2} & d_{xz} & d_{yz} & d_{x^2-y^2} & d_{xy}  \cr
    d_{z^2}   & 3.784 & 0.002 & 0 & $-$0.007 & 0 \cr
    d_{x^2-y^2}  & 0.002 & 4.583 & 0 & \textbf{$-$0.592} & 0 \cr
    d_{yz}   & 0 & 0 & 4.578 & 0 & \textbf{0.584} \cr
    d_{x^2-y^2}   &  $-$0.007 & \textbf{$-$0.592} & 0 & 4.153 & 0  \cr
    d_{xy}  & 0 & 0 & \textbf{0.584} & 0 & 4.152  \cr
      } \qquad
\end{aligned} 
\label{ham:noRot}      
\end{equation}

\begin{figure*}[!htb]
\begin{center}
\includegraphics[width=0.8\textwidth, angle=0]{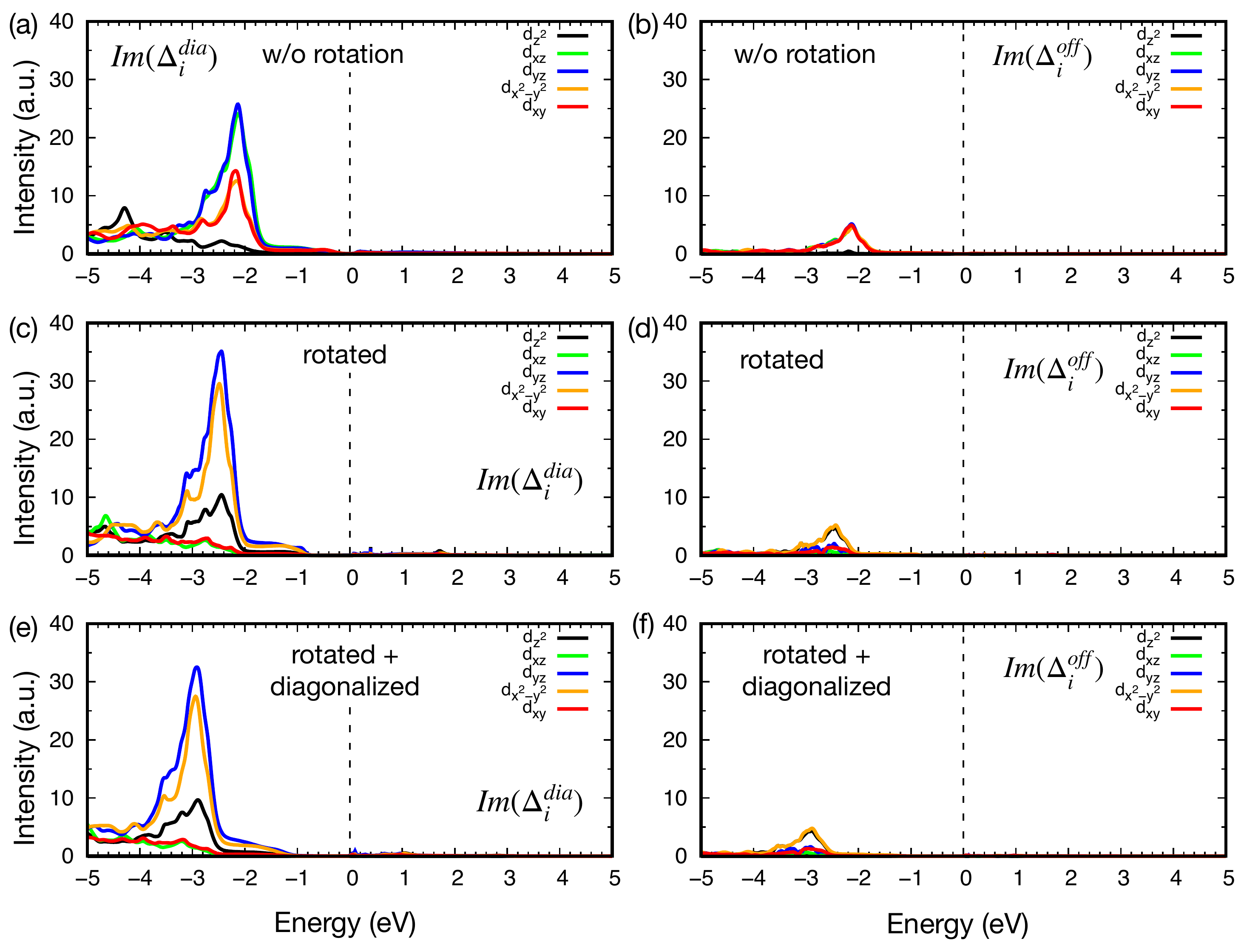}
\caption{  (a), (c), and (e) diagonal elements of the hybridization function,
and  (b), (d), and (f) off-diagonal elements of the hybridization function.
(a), (b) global coordinate is used for Wannier projection, 
(c), (d) local coordinate obtained by unitary rotation matrix is used for Wannier projection, 
and (c), (d) block diagonal part of Mn $d$ Hamiltonian is diagonalized
The calculations use $U=5$ eV,  $J=0.9$ eV, and a temperature of 300 K. 
}
\label{hyb_function}
\end{center}
\end{figure*}

Using this Hamiltonian, we employed DFT+DMFT calculations 
and plot the spectral functions in Fig. \ref{dmft_dos_pd}(a). 
Considering that the battery operates at room temperature, 
we set the temperature to 300 K. 
Since $T_N$ = 65 K of Li$_2$MnO$_3$ is much lower than 300 K, 
we focus on the paramagnetic phase. 
When using the non-rotated basis in the DMFT calculations, the resulting 
energy gap ($E_g$) is only 0.4 eV, significantly smaller than the experimental value.

To mitigate the large error of $E_g$ caused by the significant off-diagonal terms, 
we also employed the local coordinate system by using different Wannier basis
that minimize the magnitude of the off-diagonal elements.
The selection of the local coordinate system is described in detail in 
Section \ref{sec:Wann_basis}. 

\begin{equation}
\begin{aligned}
& \mathbf{H}_{rot} = \\
& \bordermatrix{ & d_{z^2} & d_{xz} & d_{yz} & d_{x^2-y^2} & d_{xy}  \cr
    d_{z^2}   & 4.069 & 0.018 & $-$0.009 & \textbf{$-$0.530} & 0.059 \cr
    d_{xz}  & 0.018 & 3.775 & 0.064 & 0.010 & 0.003 \cr
    d_{yz}   & $-$0.009 & 0.064 & 4.997 & $-$0.011 & 0.006 \cr
    d_{x^2-y^2}   &  \textbf{$-$0.530} & 0.010 & $-$0.011 & 4.668 & $-$0.101  \cr
    d_{xy}  & 0.059 & 0.003 & 0.006 & $-$0.101 & 3.754  \cr
      } \qquad
\end{aligned}      
\label{ham:rot}      
\end{equation}

As shown in Eq. \ref{ham:rot}, the off-diagonal terms are significantly reduced, 
leading to an increase in $E_g$ to 0.6 eV.
However, the obtained value of $E_g$ is still significantly smaller than the experimental gap.
As depicted in Fig. \ref{dmft_dos_pd}(b), the underestimation of $E_g$ 
can be attributed to the small value of $\Delta_{CF}$. 
$\Delta_{CF}$ within DMFT with rotated Wannier axes is around 1.5 eV, 
which is considerably smaller than the value within DFT+$U$ ($\sim$3 eV).  
In contrast, $\Delta_{dd}$ is approximately 2.7 eV and has a negligible effect on $E_g$.

To resolve the underestimation of $E_g$ within DFT+DMFT, we diagonalize 
each Mn $d$ block of the Hamiltonian (orange boxes in Fig. \ref{matrix_diag}), 
which can be obtained by the unitary rotation matrix. 
Diagonalized Wannier Hamiltonian is shown in Eq. \ref{ham:diag}.
\begin{equation}
\mathbf{H}_{dia} = 
\begin{pmatrix}
      & 3.745 & 0 & 0 & 0 & 0 \\
     & 0 & 3.788 & 0 & 0 & 0 \\
     & 0 & 0 & 4.998 & 0 & 0 \\
      &  0 & 0 & 0 & 4.985 & 0  \\
      & 0 & 0 & 0 & 0 & 3.741  \\
\end{pmatrix}    
\label{ham:diag}      
\end{equation}
%
%
As presented in Fig. \ref{dmft_dos_pd}(b), $E_g$ within DMFT is increased to 0.8 eV.
However, this value is still smaller than the experimental gap. 
In order to further investigate this discrepancy, we explore different values of 
$U$ and the double counting parameter $\alpha$
using the diagonalized Hamiltonian (see Eq. \ref{eq:Vdc}).
Surprisingly, we observe that both $\Delta_{CF}$ and the resulting $E_g$ 
are not significantly affected by variations in $U$. 
As shown in Fig. \ref{dmft_dos_pd_U}, the values of $E_g$ 
with $U$=5 and 7 are 0.8 and 0.9eV, respectively.

In contrast, we find that both $\Delta_{CF}$ and the resulting $E_g$ 
are more sensitive to $\alpha$. 
For instance, when $U$= 5 eV, the value of $E_g$ increases to 1.4 eV for $\alpha$ = 0.8. 
Furthermore, for $\alpha$ = 0.8, increasing $U$ from 5 to 7 eV further enhances 
$E_g$ from 1.4 to 2.0 eV. 
The sensitivity of the band gap $E_g$ to $\alpha$ suggests 
the importance of $p$-$d$ hybridization ($U_{pd}$) in determining $\Delta_{CF}$, 
due to the strong Mn $d$--O $p$ hybridization in Li$_2$MnO$_3$. 
It is important to note that that modifying $U_{dd}$ alone for the $pd$ model Hamiltonian 
is insufficient, as it primarily affects the $d$-$d$ correlation without adequately 
addressing the $p$-$d$ covalency.
Therefore, considering both $d$-$d$ correlation and $p$-$d$ hybridization is crucial 
for a comprehensive understanding of the electronic structure 
of Li$_2$MnO$_3$.


It should be noted that the off-diagonal elements of the local Green's 
function can remain small, regardless of the basis function.
To study this, we examine both the diagonal and off-diagonal components of 
hybridization function obtained using global coordinate projection, 
local coordinate projection, and diagonalized Wannier Hamiltonian.
The hybridization function is defined as 
\begin{equation}
\hat{\Delta} (\omega) = (\omega + \mu)\hat{I} - \epsilon_{imp} 
- \hat{\Sigma}(\omega) - \left[ \hat{G^{cor}}(\omega) \right]^{-1}
\end{equation}
where $\epsilon_{imp}$ is the matrix representing the impurity levels of correlated orbitals,
and $G^{cor}$ is correlated Green's function \cite{DMFTwDFT}. 
For each correlated atom, the hybridization function has a matrix form 
$\Delta_{ij}$, where $i$ and $j$ 
are orbital indices (5 $d$ orbitals).
We define the diagonal elements of the hybridization function as
$\Delta^{\mathrm{dia}}_{i} = |\Delta_{ii}|$ ($i$= 1 to 5), and
the off-diagonal elements are defined by 
$\Delta^{\mathrm{off}}_{i} = \sum_{j \neq i}|\Delta_{ij}|$/4.
We compare the imaginary part of the diagonal and off-diagonal element,
i.e., $Im\left(\Delta^{\mathrm{dia}}_{i}\right)$ and $Im\left(\Delta^{\mathrm{off}}_{i}\right)$.
As presented in in Figure \ref{hyb_function}, the diagonal terms are much larger than the off-diagonal terms, $\Delta^{\mathrm{off}}_{i}$.
regardless of the Wannier projection and the diagonalization of the Wannier Hamiltonian. 
The use of different basis functions leads to energy shift  
of the peak for the diagonal part of the hybridization function, 
$\Delta^{\mathrm{dia}}_{i}$.
Note that the real part of the hybridization function has similar trend; diagonal term is also dominant.

\begin{figure}
\begin{center}
\includegraphics[width=0.40\textwidth, angle=0]{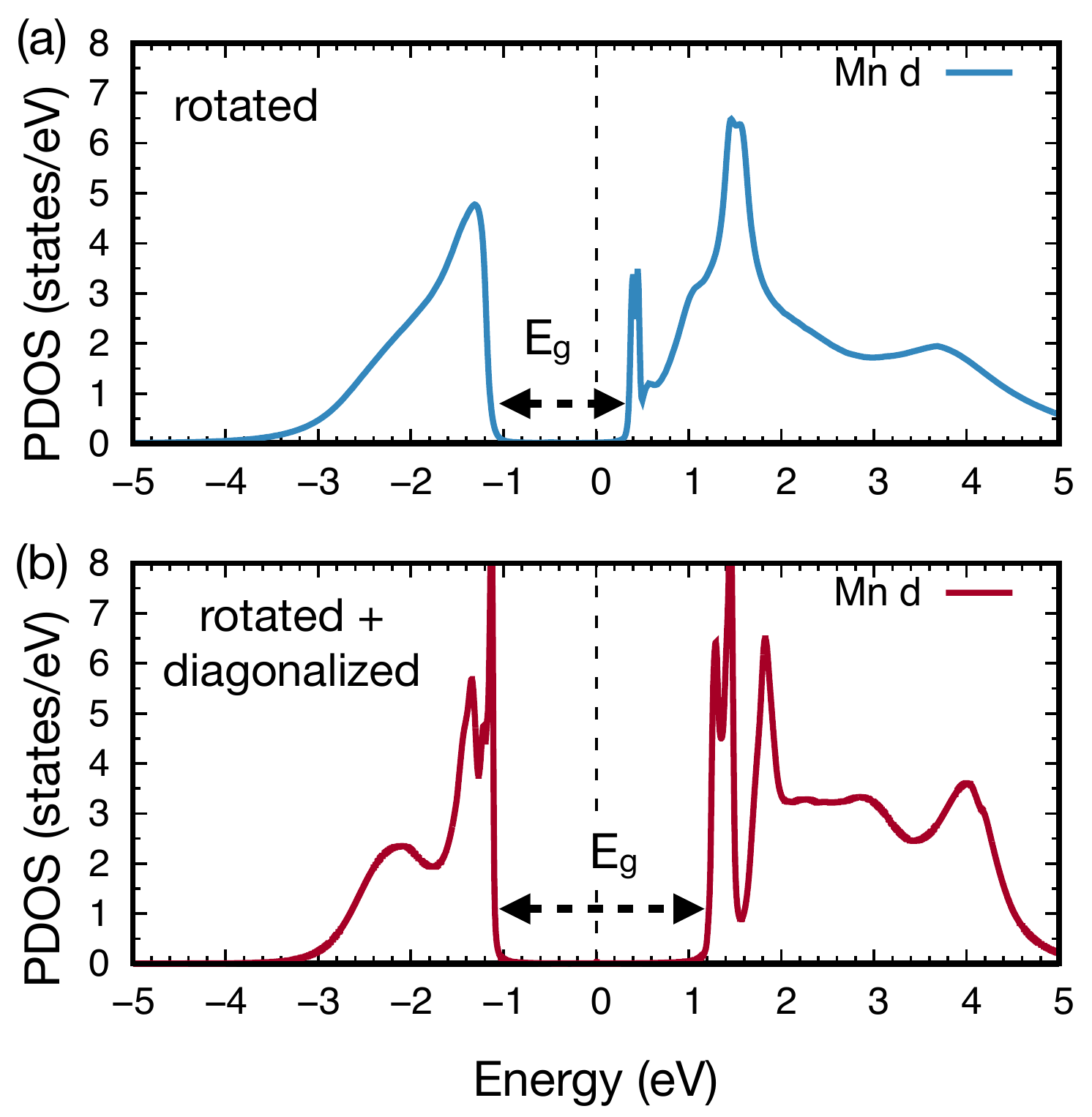}
\caption{DFT+DMFT spectral functions of Mn $d$ Wannier orbitals with $d$-only model.
(a) local coordinate obtained by unitary rotation matrix is used for Wannier projection, and 
(c) block diagonal part of Mn $d$ Hamiltonian is diagonalized.
The calculations use $U=2.0$ eV,  $J=0.9$ eV, and a temperature of 300 K. 
The length of the arrows ($E_{g}$) serves as a guide for the eyes.
}
\label{dmft_dos_d}
\end{center}
\end{figure}

Since both $d$-$d$ correlation and $p$-$d$ hybridization are important on the $E_g$,
we also consider the $d$-only model Hamiltonian as 
an alternative approach to describe the electronic structure of Li$_2$MnO$_3$. 
In the $d$-only model, the Wannier $d$ orbitals represent the hybridized 
Mn $d$ and O $p$ orbitals, in contrast to the $pd$ model where they closely 
resemble the Mn $d$ orbital. 
Therefore, the effect of $U$(Mn) on the $d$-only Wannier Hamiltonian is 
similar to the combined effects of $U_{dd}$ and $U_{pd}$ in the $p-d$ model Hamiltonian. 
As a result, both $\Delta_{CF}$ and $E_g$ are sensitive to $U$(Mn) within the $d$-only model. 
By employing Wannier projection with local coordinates, we obtain $E_g$ 
of 1.3 eV with $U$(Mn) = 2.0 eV, as shown in Fig. \ref{dmft_dos_d}(a). 
Similarly to the $pd$ basis, the diagonalization of the Wannier Hamiltonian 
further enhances $E_g$. 
In this case, the resulting $E_g$ value becomes comparable to $U$(Mn), 
yielding $E_g$ = 2.1 eV with $U$(Mn) = 2.0 eV, as presented in Fig. \ref{dmft_dos_d}(b).

\section{Summary}

In this work, we show that the off-diagonal terms of Mn $d$ Hamiltonian and $p$-$d$ 
interaction play crucial role on the electronic structure of Li$_2$MnO$_3$, 
through DFT+DMFT calculations.
Our findings underscore the necessity of accounting for the off-diagonal terms in the 
Mn $d$ block of the Wannier Hamiltonian.
The monoclinic symmetry ($C2/m$) of Li$_2$MnO$_3$ results in large 
off-diagonal terms in the Mn $d$ block when using the global coordinate, 
leading to a significant suppression of the energy gap compared to the experimental value. 
By adopting a local coordinate, the magnitude of the off-diagonal terms can be reduced, 
resulting in a substantial increase in the band gap, although it remains smaller 
than the experimental value.

To address this limitation, we diagonalize the Mn $d$ block of the Wannier Hamiltonian 
by applying a unitary rotation matrix, leading to a further enhancement of the energy gap. 
However, even with this approach, the energy gap is still not large enough.
We find that the strong hybridization between Mn $d$ and O $p$ necessitates 
considering the $p$-$d$ covalency through the adjustment of the 
double counting parameter, in addition to increasing $U_{dd}$. 
In support of this concept, we also explore a $d$-only model Hamiltonian, 
which captures the hybridized state of the Mn $d$ and O $p$ orbitals. 
Our findings demonstrate the sensitivity of the energy gap to $U$, highlighting the 
importance of both $dd$ correlation and $pd$ hybridization 
in the electronic structure of Li$_2$MnO$_3$.

These findings suggest a rigorous application of the Wannier Hamiltonian in the 
exploration of low-symmetry materials through DFT+DMFT study.
Moreover, we found that the magnetic ground state of Li$2$MnO$3$ strongly 
depends on the choice of $U$. 
While the antiferromagnetic states ($\Gamma_{2u}$ or $\Gamma_{3g}$) are more stable than 
the ferromagnetic state in the wide range of $U$ ($U\leq 5$), experimentally observed ground state 
$\Gamma_{2u}$ phase is most stable with $U\leq 2$.
Hence, careful consideration of $U$ is essential for future DFT+$U$ studies of Li$_2$MnO$_3$.

\newpage
\section{Acknowledgments}
This research is supported by the Vehicle Technologies Office
(VTO), Department of Energy (DOE), USA, through the
Battery Materials Research (BMR) program.
H. Park acknowledges financial support from the U.S. Department of Energy, Office of Science, Office of Basic Energy Sciences, Materials Science and Engineering Division.
We gratefully acknowledge the computing resources provided on Bebop, a high-performance computing cluster operated by the Laboratory Computing Resource Center at the Argonne National Laboratory.

\bibliography{myref}

	
\appendix

\section{Effect of the Coulomb interaction}
\label{appendix:coulomb}

\begin{figure*}[!htb]
\begin{center}
\includegraphics[width=0.8\textwidth, angle=0]{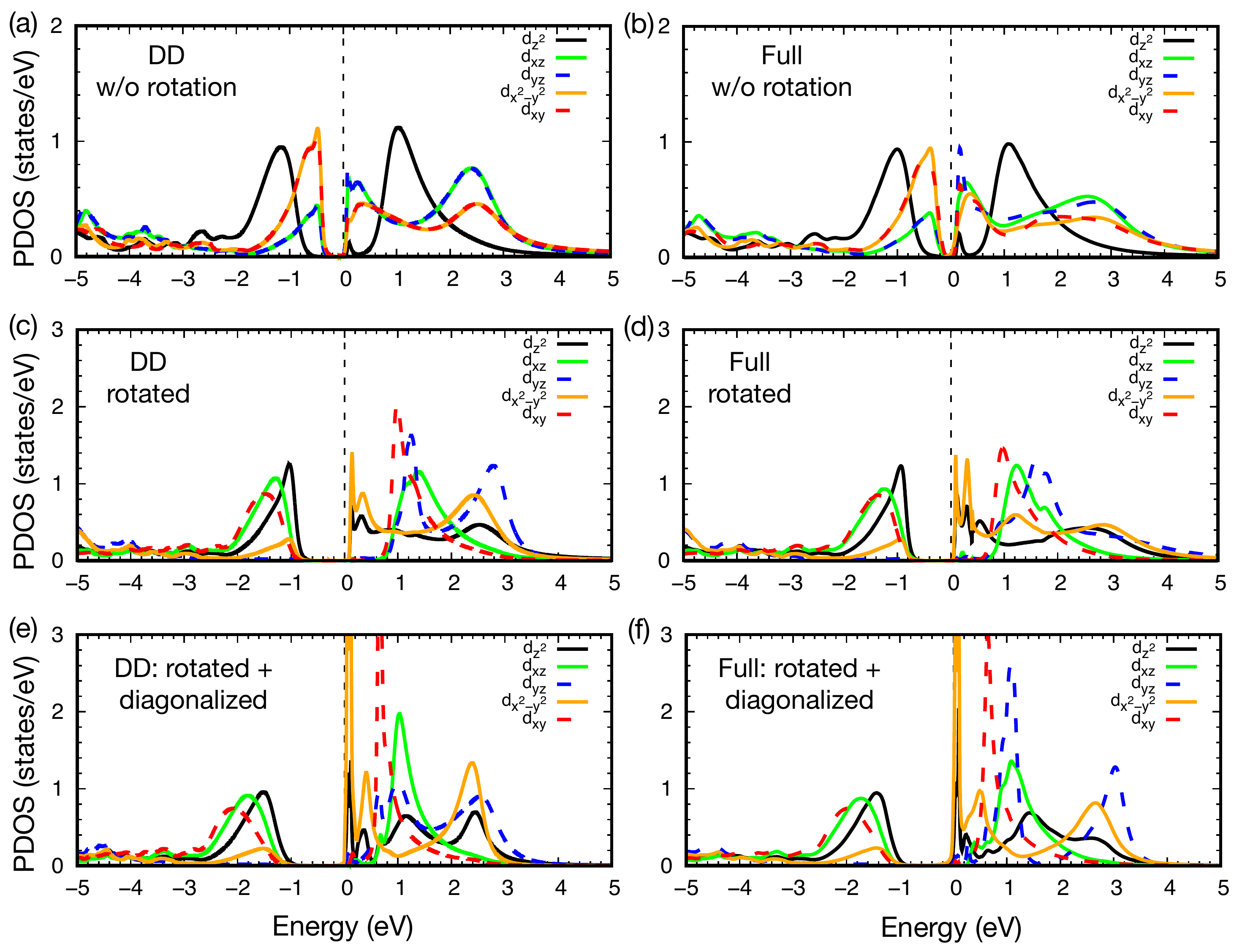}
\caption{  DFT+DMFT spectral functions of Mn $d$ Wannier orbitals, with $pd$ model.
(a), (c), and (e) results using density-density  approximation, and (b), (d), and (f) results using full Coulomb interaction.
(a), (b) global coordinate is used for Wannier projection, 
(c), (d) local coordinate obtained by unitary rotation matrix is used for Wannier projection, 
and (c), (d) block diagonal part of Mn $d$ Hamiltonian is diagonalized
The calculations use $U=5$ eV,  $J=0.9$ eV, and a temperature of 300 K. 
}
\label{full_dd}
\end{center}
\end{figure*}

In Figure \ref{full_dd} we compare the spectral functions using full Coulomb 
interaction parameterized by Slater integral, including density-density,
spin-flip, and pair-hopping interactions.
We also used only the density-density interaction term (density-density  approximation),
and compared with the full Coulomb vertex calculation.
Results are qualitatively similar, while the energy gap is reduced with full interaction
for the global coordinate Wannier projection.

\section{DFT+$U$ with nonzero $J$}
\label{appendix:dftu+J}

\begin{figure*}[!htb]
\begin{center}
\includegraphics[width=0.7\textwidth, angle=0]{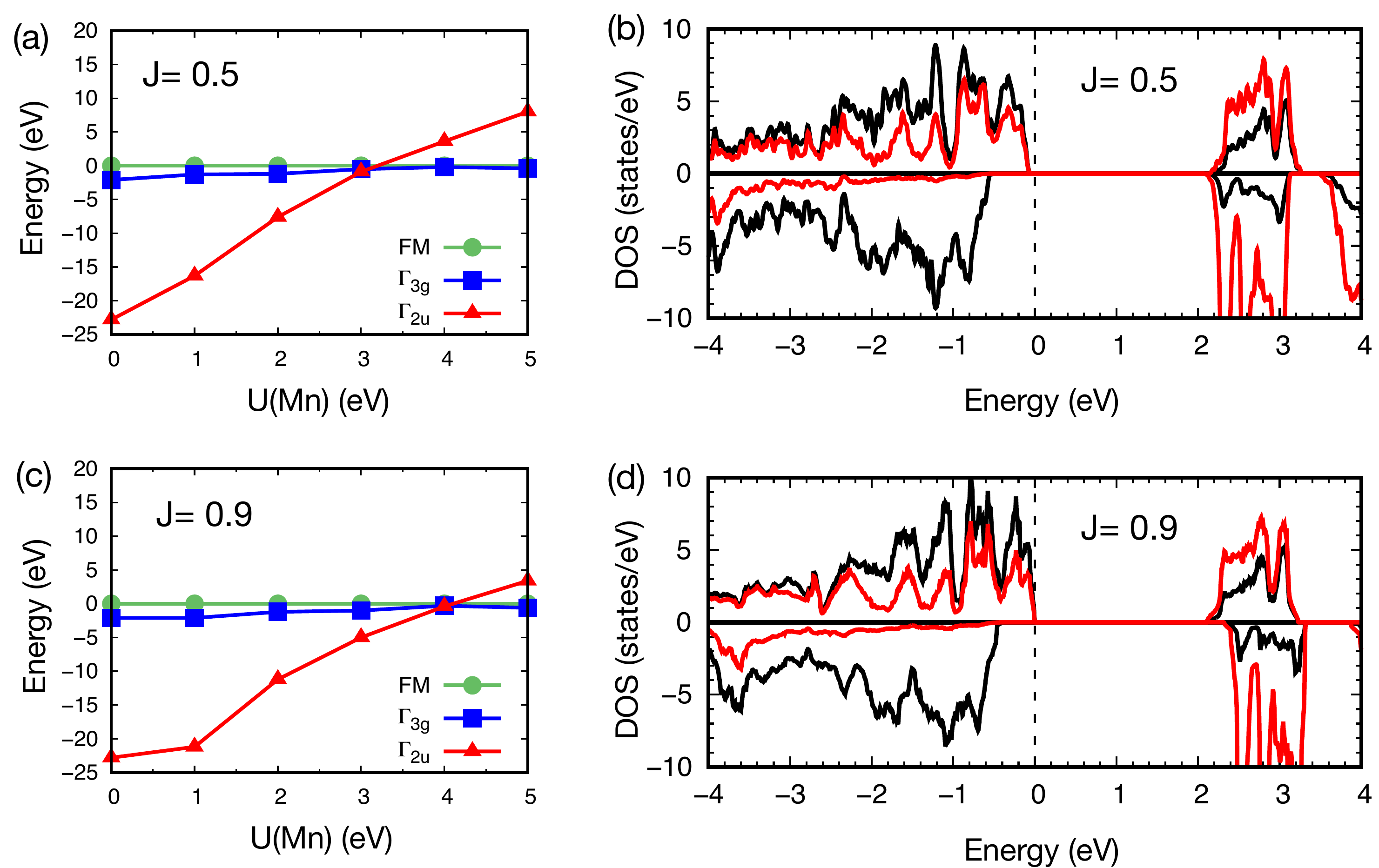}
\caption{  The relative energies of the ferromagnetic, $\Gamma_{2u}$, and $\Gamma_{3g}$ phases, 
as a function of $U$(Mn) when (a) $J$=0.5 and (c) $J$=0.9.
Energy of the ferromagnetic phase is set to be zero. 
Projected density of states (PDOS) onto the Mn $d$ (red) and 
O $p$ (black) in Li$_2$MnO$_3$ with $U$(Mn)$= 2$, using (b) $J$=0.5 and (d) $J$=0.9.
}
\label{dft_dos_sup}
\end{center}
\end{figure*}

In Figure \ref{dft_dos_sup}, we compare the electronic structures using nonzero $J$ = 0.5 and 0.9,
and also compute the relative energies of the different magnetic configurations.
We note that the electronic structures of Li$_2$MnO$_3$ with nonzero $J$ are qualitatively 
similar toe the results with $J$=0, while the critical $U$(Mn) for magnetic stability is increased
with larger $J$. 
$\Gamma_{2u}$ configuration is the ground state for $U < 3$ and $U<4$ for $J$=0.5 and 0.9,
respectively.

\end{document}